\documentclass[11pt,twoside]{article}
\usepackage{./asp2014}

\aspSuppressVolSlug
\resetcounters

\begin{document}

\title{A systematic DECam search for RR Lyrae in the outer halo of the Milky Way}
\author{Gustavo E. Medina,$^1$ Ricardo R. Muñoz,$^2$ Jeffrey L. Carlin,$^3$ A. Katherina Vivas,$^4$ Camilla J. Hansen,$^5$ Eva K. Grebel,$^1$}

\affil{$^1$Astronomisches Rechen-Institut, Zentrum f{\"u}r Astronomie der Universit{\"a}t Heidelberg, 69120 Heidelberg, Germany \\
\email{gustavo.medina@uni-heidelberg.de}}
\affil{$^2$Departamento de Astronom\'ia, Universidad de Chile, Santiago, Chile}%Camino    del    Observatorio    1515, Las Condes,
\affil{$^3$LSST, Tucson, AZ 85719, USA} 
\affil{$^4$Cerro Tololo Inter-American Observatory, NSF's National Optical-Infrared Astronomy Research Laboratory, La Serena, Chile}%Casilla 603, 
%\email{lemasle@uni-heidelberg.de}}
\affil{$^5$Max Planck Institute for Astronomy, 69117 Heidelberg, Germany} %K{\"o}nigstuhl 17, 
%\email{grebel@ari.uni-heidelberg.de}}

%\affil{$^3$Institution Name, Institution City, State/Province, Country; \email{AuthorEmail@email.edu}}}

%-----------------------------------------

% This section is for ADS Processing.  There must be one line per author.
%\paperauthor{Sample~Author1}{Author1Email@email.edu}{ORCID_Or_Blank}{Author1 Institution}{Author1 Department}{City}{State/Province}{Postal Code}{Country}
\paperauthor{Gustavo E. Medina}{gustavo.medina@uni-heidelberg.de}{0000-0003-0105-9576}{Zentrum f{\"u}r Astronomie der Universit{\"a}t Heidelberg}{Astronomisches Rechen-Institut}{Heidelberg}{}{69120}{Germany}

\paperauthor{Ricardo R. Muñoz}{rmunoz@das.uchile.cl}{}{Universidad de Chile}{Departamento de Astronom\'ia}{Santiago}{Las Condes}{}{Chile}

\paperauthor{Jeffrey L. Carlin}{jcarlin@lsst.org}{0000-0002-3936-9628}{LSST}{}{Tucson}{Arizona}{85719}{USA}

\paperauthor{A. Katherina Vivas}{kvivas@ctio.noao.edu}{0000-0003-4341-6172}{Cerro Tololo Inter-American Observatory}{}{La Serena}{}{}{Chile}

\paperauthor{Camilla J. Hansen}{hansen@mpia-hd.mpg.de}{0000-0002-7277-7922}{Max Planck Institute for Astronomy}{}{Heidelberg}{}{69117}{Germany}

\paperauthor{Eva K. Grebel}{grebel@ari.uni-heidelberg.de}{0000-0002-1891-3794}{Zentrum f{\"u}r Astronomie der Universit{\"a}t Heidelberg}{Astronomisches Rechen-Institut}{Heidelberg}{}{69120}{Germany}

%-----------------------------------------

\begin{abstract}
The discovery of very distant stars in the halo of the Milky Way provides valuable tracers on the Milky Way mass and its formation. 
Beyond $\sim$100\,kpc from the Galactic center, most of the stars are likely to be in faint dwarf galaxies or tidal debris from recently accreted dwarfs, making the outer reaches of the Galaxy important for understanding the Milky Way’s accretion history. 
However, distant stars in the halo are scarce. 
In that context, RR Lyrae are ideal probes of the distant halo as they are intrinsically bright and thus can be seen at large distances, follow well-known period-luminosity relations that enable precise distance measurements, and are easily identifiable in time-series data. 
Therefore, a detailed study of RR Lyrae will help us understand the accreted outskirts of the Milky Way.
%in more detail. 
In this contribution, we present the current state of our systematic search for distant RR Lyrae stars in the halo using the DECam imager at the 4\,m telescope on Cerro Tololo (Chile).
The total surveyed area consists of more than 110 DECam fields ($\sim$ 350\, sq. deg) and includes two recent independent campaigns carried out in 2017 and 2018 with which we have detected $>$ 650 candidate RR Lyrae stars. Here we describe the methodology followed to analyze the two latest campaigns.
Our catalog contains a considerable number of candidate RR Lyrae beyond 100\,kpc, and reaches out up to $\sim$ 250\,kpc. 
The number of distant RR Lyrae found is consistent with recent studies of the outer halo. These stars provide a set of important probes of the mass of the Milky Way, the nature of the halo, and the accretion history of the Galactic outskirts.

\end{abstract}

\section{Introduction}

Only a handful of Milky Way RR Lyrae stars have been identified at distances $>$100\,kpc from the Galactic center. These stars play an important role for disentangling the accretion history of the Galaxy given that they are not expected to have formed at these remote distances, and simulations suggest that they likely originated in recently-accreted satellite galaxies \citep{BJ2005,Sanderson2017}. Other studies have conjectured that they can also be used as potential tracers of previously unknown faint satellite systems \citep[see, e.g.,][]{BakerWillman2015}.

To reconstruct the outer halo accretion history one can, for instance, measure the shape of the Galactic density profile since these profiles have shown to be sensitive to properties such as the halo formation time, the accreted stellar mass, and how long ago the last merger occurred \citep{Pillepich2014}.
In that regard, RR Lyrae have been particularly useful for building number density counts, due to their nature as precise distance indicators \citep{Medina2018,Hernit2018}. 
Therefore, finding them at large distances plays a key role for a correct interpretation of these profiles.

Recently, \citet{Stringer2019} identified $\sim$ 6000 RR Lyrae using data from the Dark Energy Survey (DES) with a random forest classifier and taking advantage of the large footprint and depth of the survey (over 5000\,deg$^2$ and $g\sim23.5$\,mag in a single exposure). Of the RR Lyrae found by their team, $\sim$ 30\% are considered new discoveries.
In what follows, we describe the methodology we followed to detect RR Lyrae in the Galactic halo as part of the Halo Outskirts With Variable STars survey (HOWVAST), which resulted in $\sim$20 new RR Lyrae candidates found beyond 100\,kpc.

\section{Survey Design}

To carry out this work, we used data obtained with the Dark Energy Camera (DECam), which is mounted on the 4\,m telescope at Cerro Tololo Interamerican Observatory, in the context of the HOWVAST survey. The survey consists of independent campaigns that took place in the 2017B and 2018A semesters, and the fields were selected to cover diverse Galactic latitudes of the Galactic halo that have not been mapped by other deep surveys such as the DES.

In the first run we observed 16 DECam fields during four consecutive half-nights in the $r-$band, with 180\,s exposures and a cadence of one hour, whereas the second run consisted of 24 fields observed during four consecutive nights, the same exposure times and a shorter cadence ($\sim$ 40\,m). This resulted in a combined total area of $\sim$ 120\,sq. degrees of the halo mapped for the survey, and light curves made up of 20 to 30 datapoints per star.
Figure~1 shows the footprint of the HOWVAST survey, and includes the fields from the High cadence Transient Survey \citep[HiTS;][]{Forster2016} in which our group searched for distant RR Lyrae in previous studies \citep{Medina2017,Medina2018}.
Observations in the $g-$band were obtained as well, in order to provide $g-r$ colors to facilitate the process of identification of RR Lyrae stars.

\section{RR Lyrae Detection}
%\textbf{
%- Jeff, can you write a little bit about the lsst pipeline used to extract the sources?}

Beginning with images that were processed through instrument signature removal by the DECam Community Pipeline \citep{Valdes2014}, source detection and photometry were carried out using the data processing pipeline currently in development for the LSST survey \citep[][]{Bosch2019}. The LSST pipeline performs detection, aperture flux measurement, and point spread function (PSF) fitting. Throughout this work, we use PSF magnitudes, since our sources of interest are stars.

The following data analysis was performed using the computer cluster \textsc{leftraru} located at the Center for Mathematical Modeling of Universidad de Chile. 
A relative zero point was computed to determine the magnitude values for all the epochs with respect to a selected reference epoch, and these magnitudes were corrected for extinction using the recalibrated dust maps from \citet{Schlafly2011}.
Stellar objects with less than 15 datapoints in their light curves, or lacking clear indications of variability, were filtered out. %\textbf{JLC: Can we just say "lacking clear indication of variability (Delta mag < 0.2)" (and remove it from the next sentence)?}
In addition, we only examined sources with $\Delta$ mag $>$ 0.2, and $g-r$ colors similar to those of typical RR Lyrae (between -0.45 and 1.0).

For period determination we used the python package P4J, which was developed for period detection on irregularly sampled and heteroscedastic time series, and the criterion chosen to be maximized by this routine was the Cauchy-Schwarz Quadratic Mutual Information \citep{Huijse2018}.
We allowed the two best periods to be selected, requiring them to be longer than 0.2\,d, shorter than 0.95\,d, and to be considered statistically significant detections.
As a final step, we performed a visual inspection of the phased light curves making sure that only RR Lyrae-like objects were selected, based on their periods, amplitudes, and light curve shapes.
In the case of RR Lyrae candidates with more than two high probability periods, we considered the four most likely ones in the power spectrum before selecting their final period.  

\articlefigure[width=1.02\textwidth]{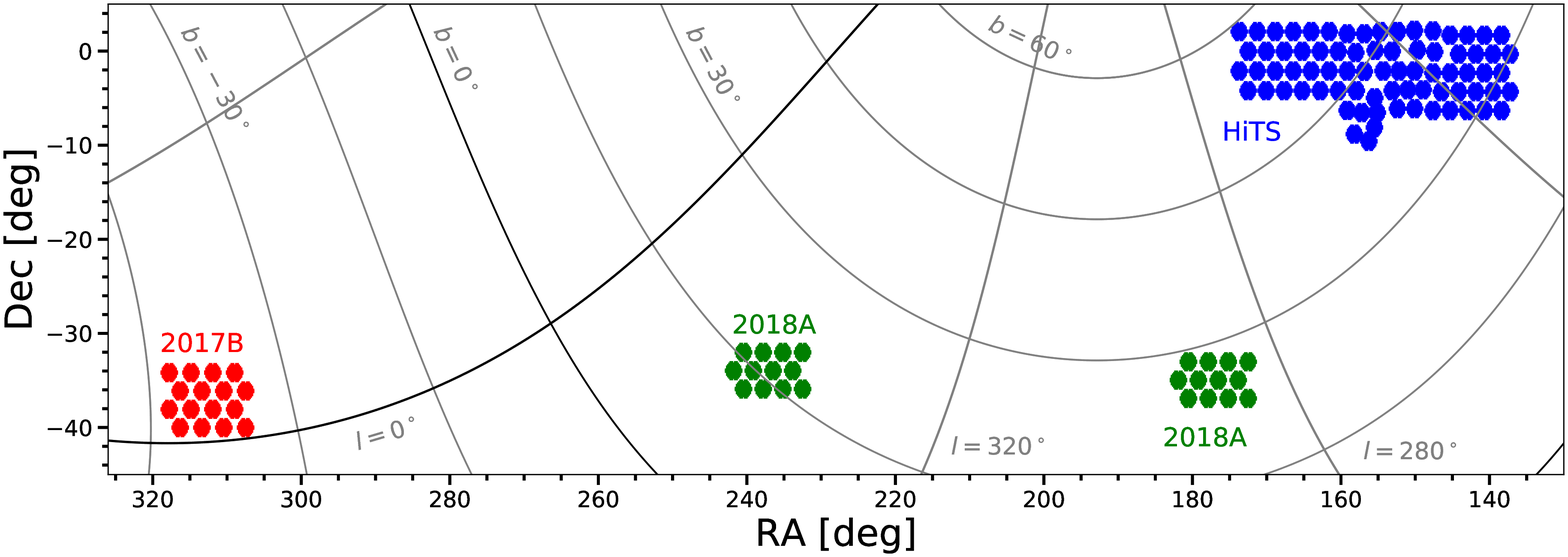}{fig1}{Spatial distribution of the DECam fields analyzed by our group so far, displayed in equatorial and Galactic coordinates. The fields observed as part of the HOWVAST survey are plotted in red and green, while the HiTS fields \citep{Forster2016} are shown in blue. In all, our deep RR Lyrae search covers $\sim350$~deg$^2$ of sky.}

\section{Early Results}

Our selection process described above resulted in 408 RR Lyrae candidates detected in the $\sim120$~deg$^2$ of 2017 and 2018 data (128 and 280 RR Lyrae, respectively). Of these, 70\% are ab-type, 29\% c-type and 1\% d-type, as depicted in the Bailey diagram shown in Figure~2. These RR Lyrae have $r$ magnitudes ranging from 15 to $\sim$22.5.
Example light curves of the detected RR Lyrae in different brightness ranges are displayed in Figure~3.
In terms of completeness, when comparing with larger RR Lyrae catalogs that overlap the surveyed regions, such as the one coming from the Catalina Real-time Transient Survey \citep{Torrealba2015,Drake2017}, we recover $\sim$ 90\% of the sources fainter than 17 in the $V$-band. 

\articlefigure[width=.55\textwidth]{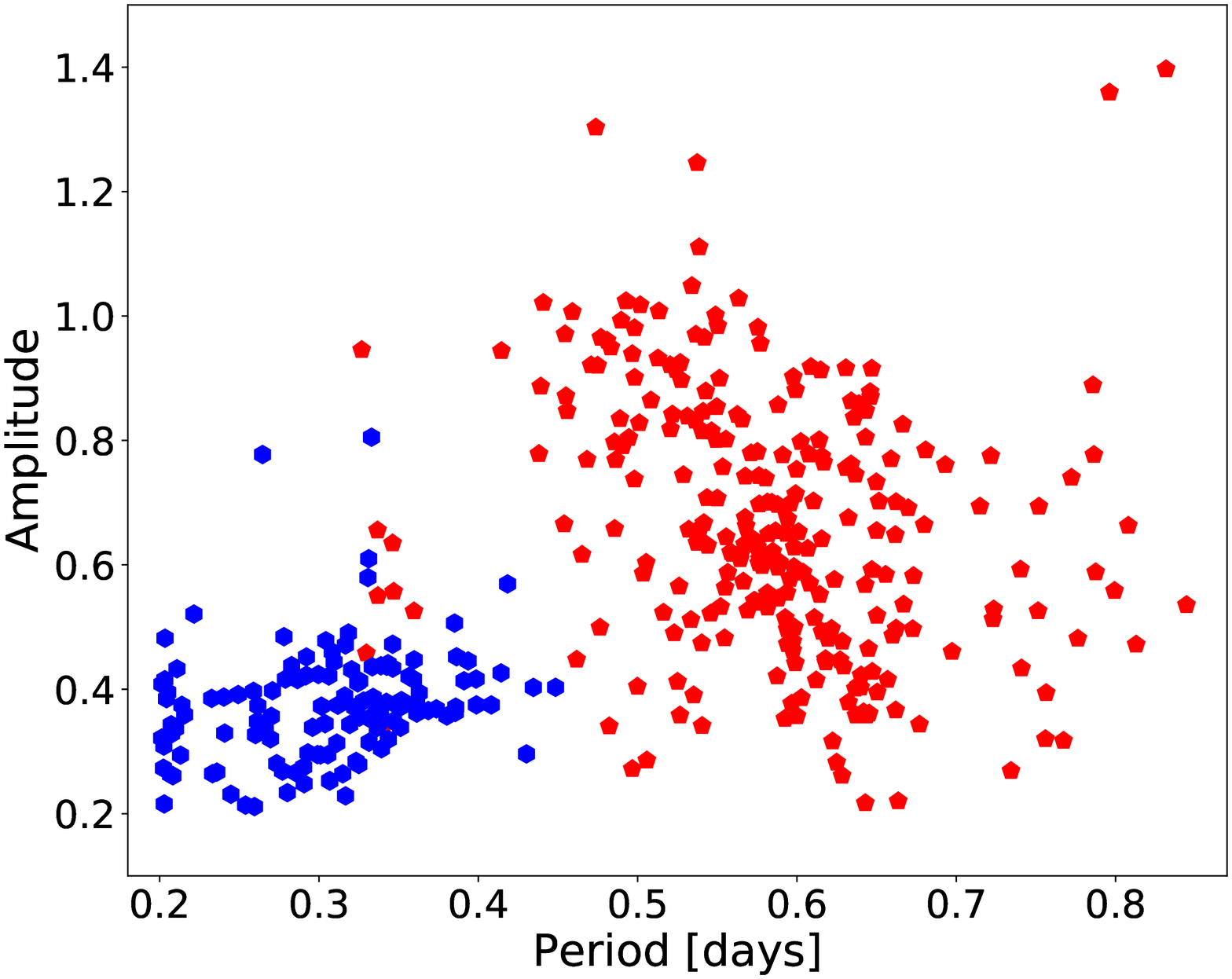}{fig2}{Bailey diagram of the sample of RR Lyrae candidates detected in this study. Blue symbols represent RRc stars, whereas RRab are displayed in red.}

To estimate the heliocentric distance to these RR Lyrae candidates, we first used the period-luminosity-metallicity (PLZ) relations computed by \citet{Sesar2017} to determine their absolute magnitudes $M_r$, assuming [Fe/H]$=-1.5$ as a representative metallicity value for halo stars. Since the PLZ relations of \citet{Sesar2017} are only suitable for ab-type RR Lyrae, we fundamentalized the period of the c-type candidates in order to estimate their absolute magnitudes, following the relationship provided by \citet{Catelan2009}.
As a result, the distances obtained range from 10 to $\sim$ 250\,kpc, and 20 candidates were found to have heliocentric distances larger than 100\,kpc.

With the estimated distances of our sample of RR Lyrae candidates, we studied the space density distribution of both ab$-$ and c$-$type RR Lyrae stars in the halo, that is, the Galactocentric distance dependence of the number of RR Lyrae stars per unit volume.
To do this, we adopted a spherical halo model and used the Markov Chain Monte Carlo routine \textsc{emcee} \citep{Foreman2013} to find the parameter distribution of the model.  
We also considered a break in the density profile as a free parameter.
%This resulted in density plots with inner slopes of $-1.35_{-0.61}^{+0.51}$ (2017), $-0.62_{-2.07}^{+1.36}$ (2018, low latitude) and $-1.79_{-0.27}^{+0.28}$ (2018, high latitude), outer slopes of $-4.95_{-0.64}^{+0.50}$ (2017), $-4.63_{-0.72}^{+0.45}$ (2018, low latitude) and $-4.68_{-0.33}^{+0.27}$ (high latitude). 
%A break of the profile was found in the three cases to be around $\sim$ 20\,kpc.
As an overall result, taking into account the distributions obtained in \citet{Medina2018} and the stars obtained in this work (under these assumptions, and not considering known structures or dwarf galaxies in the profiles), the inner and outer halo slope are found to be $-1.06_{-0.84}^{+0.25}$ and $-4.36_{-0.11}^{+0.10}$ respectively, with a break radius at 20\,kpc.

\articlefiguretwo{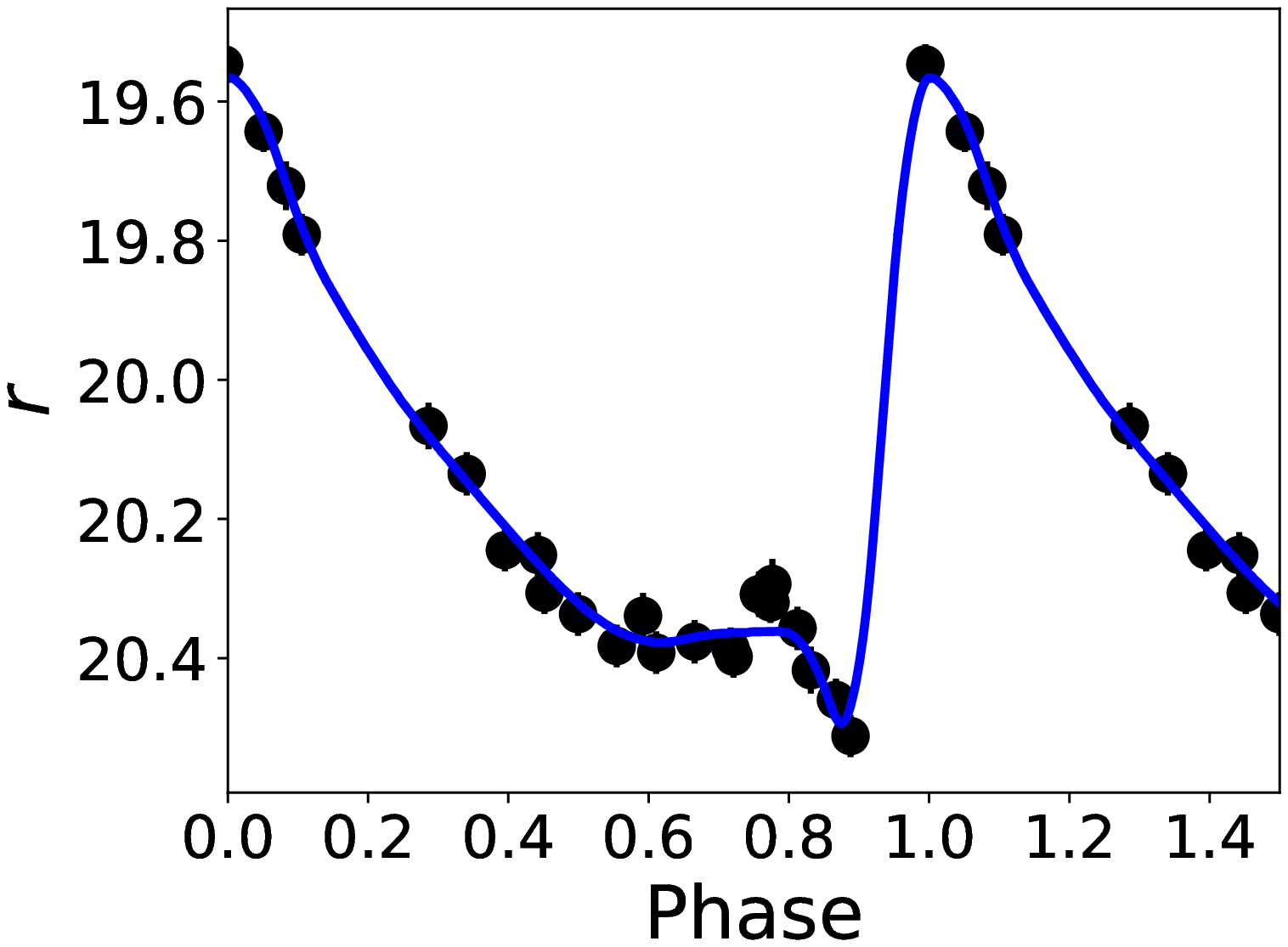}{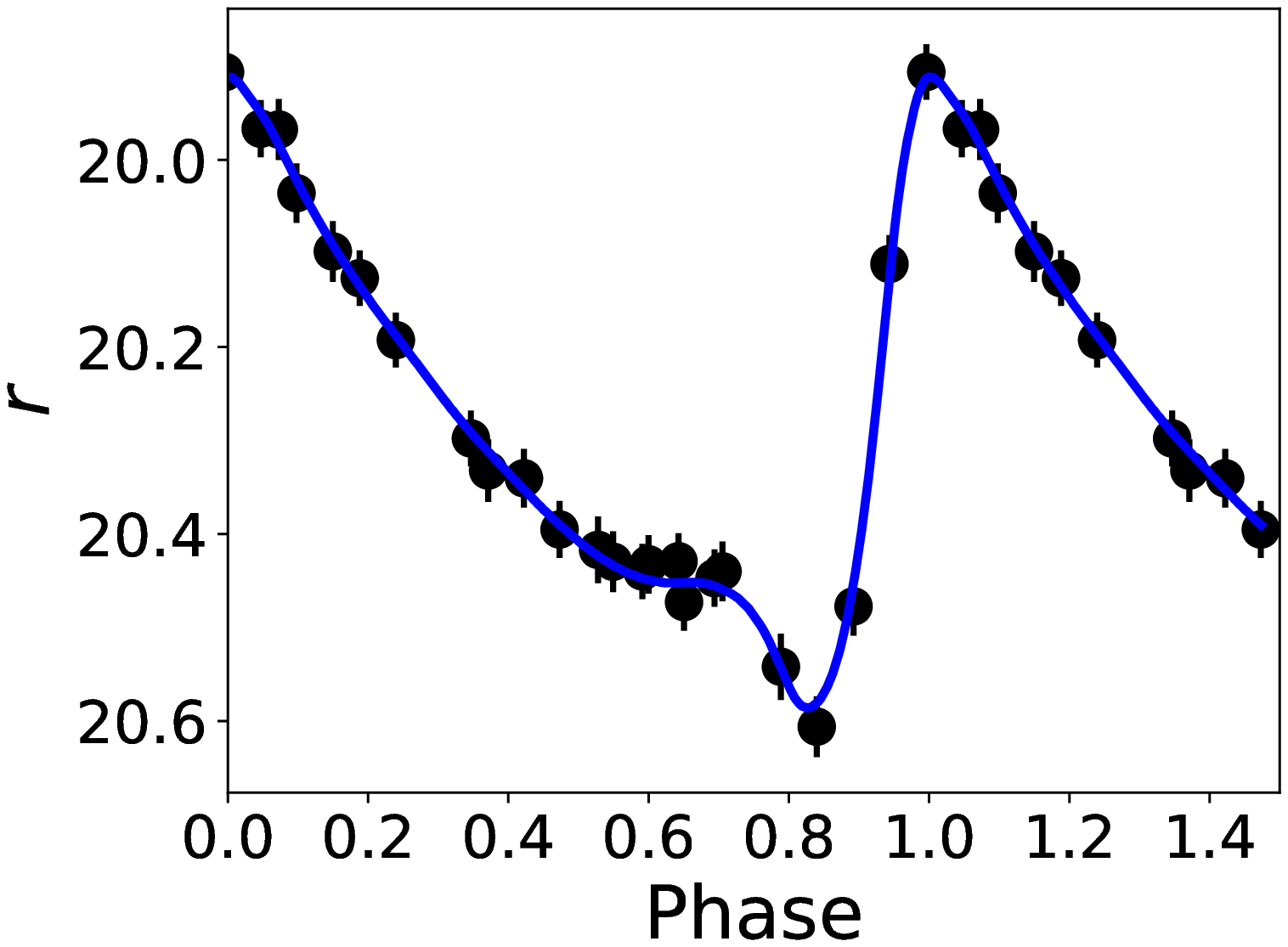}{fig3}{Examples of phased light curves from our catalog. 
The templates used were obtained through the code \textsc{gatspy} \citep{VanderPlas2015}, which uses light curve templates based on SDSS Stripe 82 RR Lyrae stars \citep{Sesar2010}.}

\section{Final Remarks}

By using data from the Dark Energy Camera, we are conducting a study to uncover the nature of the outskirts of the Galaxy, and its formation, by investigating the population of remote Milky Way RR Lyrae. 
%We have designed HOWVAST as an effort to map the outer limits of the Galaxy, and as a result of the early analysis of its data we have found, in addition to what we presented in our previous study \citep{Medina2017,Medina2018}, more than 650 RR Lyrae stars in an area that covers $\sim$ 350\,sq. degrees (408 RR Lyrae in the HOWVAST fields only).
As a continuation of our previous study (using HiTS data)%\citep[using HiTS data;][]{Medina2017,Medina2018}
, we designed the HOWVAST survey as an effort to specifically map the outer limits of the Galaxy.
Using HOWVAST data we identified $\sim$400 RR Lyrae star candidates in $\sim120$~deg$^2$, which when combined with the RR Lyrae in DECam HiTS data analyzed by our team \citep{Medina2017, Medina2018} adds up to $>$ 650 RR Lyrae stars in $\sim350$~deg$^2$ of sky.
%As a result of the early analysis of HOWVAST data we have found 408 additional RRLyrae in these fields alone.

With these stars we have constructed preliminary number density profiles, and found that the results are consistent with simulations that build up the halo using accretion-only models, with which a reasonable value for the outer halo slope is -4.4 \citep{Deason2013}. Other power law indices from the literature are summarized in Table 4 from our previous work \citep{Medina2018}.

Finally, the total number of remote RR Lyrae identified in this work is in line with the results shown by \citet{Stringer2019} using DES data. 
It is worth noting that, even though our findings are still below the number of distant RR Lyrae expected from theoretical predictions \citep{Sanderson2017}, these programs form part of
an important group of surveys that are precursors for future deep-large sky programs that will provide a more complete vision of the halo populations.%, such as the Large Synoptic Survey Telescope.

\acknowledgements 
Powered@NLHPC: This research was partially supported by the supercomputing infrastructure of the NLHPC (ECM-02). This project used data obtained with the Dark Energy Camera, which was constructed by the Dark Energy Survey collaboration. GM acknowledges the support of the Hector Fellow Academy.

\end{document}